\documentclass{article}
\usepackage{amsmath}
\usepackage{amssymb}
 \newcommand{\begeq}{\begin{equation}}
\newcommand{\bea}{\begin{eqnarray}}
\newcommand{\eea}{\end{eqnarray}} \newcommand{\nn}{\nonumber}

\newcommand{\ca}{$C^*$-algebra} 
 \newcommand{\rep}{representation}

\newcommand{\Hs}{Hilbert space} 
 
\newcommand{\sta}{$\mbox{}^*$-algebra}

   \newcommand{\vna}{von
Neumann algebra}
\newcommand{\op}{^{\mbox{\tiny op}}}
\newcommand{\id}{\mbox{\rm id}}

 \newcommand{\ovl}{\overline}
 \newcommand{\til}{\tilde}
\newcommand{\raw}{\rightarrow} 
\newcommand{\rac}{\rightarrowtail}
\newcommand{\lac}{\leftarrowtail}
\newcommand{\law}{\leftarrow}

\newcommand{\rlh}{\rightleftharpoons} 
\newcommand{\ot}{\otimes} 
\newcommand{\la}{\langle} \newcommand{\ra}{\rangle}

\newcommand{\x}{\times} 
 
\newcommand{\cin}{C^{\infty}} \newcommand{\cci}{C^{\infty}_c}

\newcommand{\inv}{^{-1}}

\newcommand{\al}{\alpha} \newcommand{\bt}{\beta}
\newcommand{\gm}{\gamma} \newcommand{\Gm}{\Gamma}
 \newcommand{\Dl}{\Delta}
 
 \newcommand{\et}{\eta}

\newcommand{\lm}{\lambda} 
\newcommand{\rh}{\rho} \newcommand{\sg}{\sigma}
 \newcommand{\ta}{\tau} 
\newcommand{\Ph}{\Phi} \newcommand{\phv}{\varphi}
\newcommand{\ch}{\chi} \newcommand{\ps}{\psi} \newcommand{\Ps}{\Psi}
\newcommand{\om}{\omega} 

\newcommand{\A}{\mathfrak{A}} \newcommand{\B}{\mathfrak{B}}
\newcommand{\GC}{\mathfrak{C}}

\newcommand{\GM}{\mathfrak{M}} \newcommand{\GN}{\mathfrak{N}}

\newcommand{\GP}{\mathfrak{P}} 

 \newcommand{\CF}{{\mathcal F}}
 \newcommand{\CE}{{\mathcal E}}
 \renewcommand{\H}{{\mathcal H}}
 
\newcommand{\CK}{{\mathcal K}}   \newcommand{\CL}{{\mathcal L}}

\newcommand{\C}{{\mathbb C}} 
 
 \newcommand{\R}{{\mathbb R}}
 
  \makeatletter
\newskip\tempskip \def\endproof{{\parfillskip24\p@ plus\@ne
fil\@@par}\tempskip\prevdepth
\ifdim\lastskip=\z@\tempskip\z@\else\vskip-\lastskip
\ifdim\tempskip>4\p@ \tempskip.5\tempskip \else \tempskip\z@\fi\fi
\nobreak\vskip-\baselineskip\vskip-\tempskip\noindent\hbox
to\hsize{\hfill
$\blacksquare$}\par\vskip\tempskip\vskip\abovedisplayskip\@doendpe}
\makeatother \makeatletter
\newskip\tempskip \def\endiproof{{\parfillskip24\p@ plus\@ne
fil\@@par}\tempskip\prevdepth
\ifdim\lastskip=\z@\tempskip\z@\else\vskip-\lastskip
\ifdim\tempskip>4\p@ \tempskip.5\tempskip \else \tempskip\z@\fi\fi
\nobreak\vskip-\baselineskip\vskip-\tempskip\noindent\hbox
to\hsize{\hfill
$\Box$}\par\vskip\tempskip\vskip\abovedisplayskip\@doendpe}
\makeatother \newcommand{\enp}{\endproof}

\newcommand{\otq}{\hat{\otimes}}
\newcommand{\otc}{\circledcirc}
\newcommand{\otg}{\circledast}

\newcommand{\Ca}{\mbox{\rm \textsf{C}$\mbox{}^*$}}
\newcommand{\Wa}{\mbox{\rm \textsf{W}$\mbox{}^*$}}
\newcommand{\Gr}{\mbox{\rm \textsf{G}}}

\newcommand{\LG}{\mbox{\rm \textsf{LG}}}
\newcommand{\MG}{\mbox{\rm \textsf{MG}}}
\newcommand{\LGc}{\mbox{\rm \textsf{LGc}}}

\newcommand{\Po}{\mbox{\rm \textsf{Poisson}}}

\newcommand{\LGb}{\mbox{\rm \textsf{LG'}}}
\newtheorem{theorem}{Theorem}[section]
\newtheorem{lemma}{Lemma}[section]
\newtheorem{definition}{Definition}[section]

\newcommand{\eqnul}{\setcounter{equation}{0}}

\begin{document}
\pagestyle{plain}
\title{Functoriality and Morita equivalence of operator algebras and
Poisson manifolds associated to groupoids\thanks{Dedicated to Dai
Evans at his 50th birthday. The results in this paper were first
presented in Cardiff on 10-10-2000.}}
\author{N.P. Landsman\thanks{Korteweg--de Vries Institute for Mathematics,
University of Amsterdam,
Plantage Muidergracht 24,
NL-1018 TV AMSTERDAM, THE NETHERLANDS, 
email \texttt{npl@science.uva.nl}}
\thanks{Supported by a Fellowship from the Royal Netherlands Academy
of Arts and Sciences (KNAW)}}
\date{\today}
\maketitle
\textbf{Abstract} 
It is well known that a measured groupoid $G$ defines a \vna\
$W^*(G)$, and that a Lie groupoid $G$ canonically defines both a
$C^*$-algebra $C^*(G)$ and a Poisson manifold $A^*(G)$.  We show that
the maps $G\mapsto W^*(G)$, $G\mapsto C^*(G)$ and $G\mapsto A^*(G)$
are functorial with respect to suitable categories.  
In these categories Morita equivalence is isomorphism of
objects, so that these maps preserve 
Morita equivalence.
\section{Introduction}
Kontsevich has introduced the idea of the ``three worlds'', viz.\
commutative, Lie, and associative algebras, relating these worlds to
each other and to ``formal'' noncommutative geometry \cite{Kon}.  In
the context of noncommutative geometry in the sense of Connes
\cite{Con}, and in particular of its relationship with quantum theory
and quantization, three other worlds are relevant, namely \vna s, \ca
s, and Poisson manifolds.  Groupoids provide access to each of these.

Firstly, measured groupoids $G$ \cite{Mackey,Ram1,Hah1,FHM,Con0,MoS}
define \vna s $W^*(G)$ in standard form \cite{CT,Hah2,Sam,Sau2,Ren}.  
Secondly, Lie groupoids $G$ \cite{Mac} canonically define
\ca s $C^*(G)$ \cite{Con1,Con}.  Thirdly, one may
canonically associate a Poisson manifold $A^*(G)$ with a Lie groupoid
$G$ \cite{CDW,Cou,DS}.

For the most basic examples of these associations, first note that a
set $S$ defines two entirely different groupoids.  The first has $S$
as the total space $G_1$, and also as the base space $G_0$ of $G$. If
$S$ is an analytic measure space $(X,\mu)$, this leads to $W^*(X)\cong
L^{\infty}(X,\mu)$, and if $S$ is a manifold $M$ one obtains
$C^*(M)\cong C_0(M)$, and $A^*(M)\cong M$ with zero Poisson bracket.
The second is the pair groupoid of $S$, with $G_1=S\x S$ and
$G_0=S$. In that case one has $W^*(X\x X)\cong \B(L^2(X,\mu))$,
$C^*(M\x M)\cong \mathfrak{K}(L^2(M))$, and $A^*(M\x M)\cong T^*(M)$.

If the groupoid is a group, one recovers the usual \vna\ and \ca\ defined
by a locally compact group. The Poisson manifold defined by a Lie group
is just the dual of the Lie algebra, equipped with the Lie--Poisson structure.
Group actions define the associated action groupoids \cite{Mac}, which
in turn reproduce the group measure space construction of Murray and
von Neumann, the notion of a transformation group \ca, and the class
of semidirect Poisson structures, respectively (for the latter cf.\
\cite{KriMar}).  For example, in the ergodic case all hyperfinite
factors arise in this way.  

Finally, the \vna s and \ca s defined by
foliations \cite{Con0,Con1,Con,MoS} may be seen as special cases of
the above constructions as well, where $G$ is the holonomy groupoid of
a smooth foliation. This class of examples formed a major motivation
for the development of noncommutative geometry.  

For fixed $G$, there are certain relationships between these
constructions.  Under appropriate technical conditions, both measured
and Lie groupoids may be seen as special instances of locally compact
groupoids with Haar system \cite{Ren}; see \cite{Ram2} and
\cite{LR,NPL3}, respectively.  The \vna\ $W^*(G)$ is then simply the
weak closure of $C^*(G)$ in its regular \rep.  The connection between
$A^*(G)$ and $C^*(G)$ is deeper: $C^*(G)$ is a strict deformation
quantization of $A^*(G) $\cite{NPL3,NPLCMP,LR}. This means, among other
things, that there exists a continuous field of \ca s over $[0,1]$,
whose fiber above 0 is the commutative \ca\ $C_0(A^*(G))$, all other
fibers being $C^*(G)$.  The \ca\ of continuous cross-sections of this
continuous field turns out to be the \ca\ of the normal groupoid
\cite{HS} defined by the embedding $G_0\hookrightarrow G_1$ of the
unit space of $G$ into its total space (Connes's tangent groupoid
\cite{Con} corresponds to the special case of a pair groupoid $G=M\x M$).

In the present paper, we examine and compare the properties of the
associations $G\mapsto W^*(G)$, $G\mapsto C^*(G)$ and $G\mapsto
A^*(G)$ as a function of $G$. Our main result is that each of these
maps is functorial, though not with respect to the obvious arrows
defining the pertinent categories.  The categories that are involved
have the desirable property that isomorphism of objects is the same as
Morita equivalence (as previously defined by Rieffel for \vna s and
\ca s \cite{Rie2} and by Xu for Poisson manifolds \cite{X2}), so that
functoriality implies that Morita equivalence is preserved.  

Often involving different terminology, for \vna s many special cases of
the latter property have been known for some time, starting with
Mackey's ergodic imprimitivity theorem \cite{Mackey,Ram1}, and
including results in \cite{FHM,Krieger,Ram2,TS}.  For \ca s and
Poisson manifolds the preservation of Morita equivalence was already
known in full generality; see \cite{MRW} and \cite{NPLvE},
respectively.  Special cases of our functoriality results include also
\cite{Mrc1,Mrc2,SO,Sta}. We surmise that the computations in \cite{HS},
taking place in the category KK of separable \ca s as objects and
KK-groups as arrows, can be generalized to arbitrary Lie groupoids;
they should then be related to our results as well.

The plan of this paper is as follows. In section 2 we deal with
measured groupoids and \vna s, in section 3 we treat Lie groupoids
and \ca s, and in section 4 we end with Lie groupoids and Poisson
manifolds.  Our main results are Theorems \ref{T1}, \ref{Thm1}, and
\ref{Thm2}. 

The reader will notice that the
category of measured groupoids and the category of Lie groupoids are defined in an apparently
totally different way. The fact that these categories are actually closely
related is explained in \cite{NPLO}, to which we refer in general for
motivation and for more details about the categories we use here.
This includes the proof  that, as already mentioned,  in each case 
recognized notions of Morita equivalence turn out to coincide
with isomorphism of objects in the pertinent category.

\textbf{Notation}
We use the notation $$
A\times^{f,g}_B C= \{(a,c)\in A\x C\mid f(a)=g(c)\}
$$
for the fiber product
of sets $A$ and $C$ with respect to maps $f:A\raw B$ and $g:C\raw B$.
The total space of a groupoid $G$ is denoted by $G_1$, and its base space
by $G_0$. The source and target projections are called
$s:G_1\raw G_0$ and $t:G_1\raw G_0$, and multiplication is a map
$m:G_2\raw G_1$, with $G_2=G_1 \x_{G_0}^{s,t}G_1$. The inversion is
denoted by $I:G_1\raw G_1$. A functor $\Ph:G\raw H$ decomposes 
into $\Ph_i:G_i\raw H_i$, $i=1,2$, subject to the usual axioms.
\section{Functoriality of $G\mapsto W^*(G)$}
\subsection{The category \MG\ of measured groupoids and  functors}
The concept of a measured groupoid emerged from the work of Mackey on
ergodic theory and group \rep s \cite{Mackey}. For the technical
development of this concept see \cite{Ram1,Hah1,FHM}. A different
approach was initiated by Connes \cite{Con0}. The connection between
measured groupoids and locally compact groupoids is laid out in
\cite{Ren,Ram2}. 
\begin{definition}\label{defmg}
A Borel groupoid is a groupoid $G$ for which $G_1$ is an
analytic Borel space,  $I$ is a Borel map,
$G_2\subset G_1\x G_1$ is a Borel subset, and multiplication
$m$ is a Borel map.  It follows that 
$G_0$ is a Borel set in $G_1$, and that $s$ and $t$ are Borel maps.

A left Haar system on a Borel groupoid is a family of measures
$\{\nu^u\}_{u\in G_0}$, where $\nu^u$ is supported on 
the $t$-fiber $G^u=t\inv(u)$, which is left-invariant in that
\begin{equation}
\int d\nu^{s(x)}(y)\, f(xy)=\int d\nu^{t(x)}(y)\, f(y) \label{LHM}
\end{equation}
for all $x\in G_1$ and all positive 
Borel functions $f$ on $G_1$ for which both
sides are finite.

A measured groupoid is a Borel groupoid equipped with a Haar system
as well as a Borel measure $\til{\nu}$ on $G_0$ with the property that
the measure class of the measure $\nu$ on $G_1$, defined by
\begin{equation}
\nu=\int_{G_0} d\til{\nu}(u)\, \nu^u, \label{mg1}
\end{equation}
is invariant under $I$ (in other words, $\nu\inv=I(\nu)\sim \nu$).
\end{definition}

Recall that the push-forward of a measure under a Borel map
is given by $t(\nu)(E)=\nu(t\inv(E))$ for Borel sets $E\subset G_0$.

This definition turns out to be best suited for categorical
considerations. It differs from the one in \cite{Ram1,Hah1}, which is
stated in terms of measure classes. However, the measure class of
$\nu$ defines a measured groupoid in the sense of \cite{Ram1,Hah1},
and, conversely, the latter is also a measured groupoid according to
Definition \ref{defmg} provided one removes a suitable null set from
$G_0$, as well as the corresponding arrows in $G_1$; cf.\ Thm.\ 3.7 in
\cite{Hah1}. Similarly, Definition \ref{defmg} leads to a locally
compact groupoid with Haar system \cite{Ren} after removal of such a
set; see Thm.\ 4.1 in \cite{Ram2}.  A measured groupoid according to
Connes \cite{Con0} satisfies Definition \ref{defmg} as well, with
$\til{\nu}$ constructed from the Haar system and a transverse measure
\cite{MoS}. See all these references for extensive information and examples.

The fact that a specific choice of a measure in its class is made in
Definition \ref{defmg} is  balanced by the concept of a
measured functor between measured groupoids, which is entirely
concerned with measure classes rather than individual measures.
Moreover, one merely uses the measure class of $\til{\nu}$.

The measure $\til{\nu}$ on $G_0$ induces a measure $\hat{\nu}$ on
$G_0/G$, as the push-forward of $\til{\nu}$ under the canonical
projection.  Similarly for a measured groupoid $H$, for whose measures
we will use the symbol $\lm$ instead of $\nu$.  
We say that a functor $\Ph$ is Borel if both $\Ph_0$ and $\Ph_1$ are. If so, $\Ph_0$ 
induces a Borel map $\hat{\Ph}_0:G_0/G\raw H_0/H$ in the obvious way.
\begin{definition}\label{defmf}
A measured functor $\Ph:G\raw H$ between two measured groupoids is
a Borel map that is algebraically a functor and satisfies
$\hat{\Ph}_0(\hat{\nu})\prec\prec \hat{\lm}$.
\end{definition}

What we here call a measured functor is called a strict homomorphism
in \cite{Ram1}, and a homomorphism in \cite{Ram2}.  Also, note that in
\cite{Mackey,Ram1,FHM} various more liberal definitions are used (in
that one does not impose that $\Ph$ be a functor algebraically at all
points), but it is shown in \cite{Ram2} that if one passes to natural
isomorphism classes, this greater liberty gains little.
\begin{definition}\label{defMG}
The category \MG\ has measured groupoids as objects, and isomorphism
classes of measured functors as arrows.  (Here a natural
transformation $\nu:G_0\raw H_1$ between Borel functors from $G$ to
$H$ is required to be a Borel map.)  Composition is defined by
$[\Ps]\circ [\Ph]=[\Ps\circ \Ph]$, and the unit arrow at a groupoid
$G$ is $1_G=[\id_G]$, where $\id_G:G\raw G$ is the identity functor.
\end{definition}
\subsection{The category \Wa\ of \vna s and correspondences}
Let $\GM,\GN$ be \vna s. Recall that an $\GM$-$\GN$ correspondence
$\GM\rac\H\lac\GN$ is given by a \Hs\ $\H$ carrying commuting normal
unital \rep s of $\GM$ and $\GN\op$.  See \cite{Con}. The notion of isomorphism of
correspondences is the obvious one: one requires a unitary isomorphism
between the Hilbert spaces in question that intertwines the left and
right actions.

Given two matched correspondences $\GM\rac\H\lac\GN$ and
$\GN\rac\CK\lac\GP$, one may define an $\GM$-$\GP$ correspondence
$\GM\rac\H\boxtimes_{\GN}\CK\lac\GP$, called the relative tensor
product or ``Connes fusion'' of the given correspondences.  This
constructions is a \vna ic version of the bimodule tensor product in
pure algebra. Various definitions exist \cite{Con,Sau,Was}, which
 coincide up to isomorphism. This composition is associative 
up to isomorphism.
A standard \rep\ of a \vna\ $\GM$ on $\H=L^2(\GM)$, unique up to
unitary equivalence, 
is best seen as an $\GM$-$\GM$ correspondence with special properties.
One of these is that $L^2(\GM)$ acts as a two-sided unit for $\boxtimes_{\GM}$,
again merely up to isomorphism.
\begin{definition}\label{defWa}
The category \Wa\ has \vna s as objects, and isomorphism classes of
correspondences as arrows, composed by the relative tensor product,
for which the standard forms $L^2(\GM)$ are units.
\end{definition}

To detail, we here regard an isomorphism class $[\GM\rac\H\lac\GN]$ as
an arrow from $\GM$ to $\GN$, so that the composition is $$
[\GN\rac\CK\lac\GP]\circ
[\GM\rac\H\lac\GN]=[\GM\rac\H\boxtimes_{\GN}\CK\lac\GP].  $$ Using
results in \cite{Rie2,Sau}, it is easily seen that two \vna s are
Morita equivalent iff they are isomorphic in \Wa\ \cite{NPLO}, and this
is true iff there is a correspondence in which the commutant of one is
isomorphic to the opposite algebra of the other, or iff they are
stably isomorphic.
\subsection{The map $G\mapsto W^*(G)$ as a functor}
It is well known that a measured groupoid defines a \vna\ in standard
form \cite{CT,Hah2,Sau2,Sam,Ren}.  In this section, we extend the map
$G\mapsto W^*(G)$ to a map from \MG\ to \Wa, and establish its
functoriality. The precise classes of Borel function $f,g$ on $G_1$ for which
the formulae below are well defined are spelled out in the above papers; for example,
one may assume that $f,g \in II(G_1)$ as defined in \cite{Hah2}.

Let $G$ be a measured groupoid (cf.\ Definition \ref{defmg}). 
Convolution on $G$ is defined by
\begeq
f*g(x)=\int_{G_1} d\nu^{s(x)}(y)\, f(xy)g(y\inv), \label{defconv}
\end{equation}
and involution is
\begeq
f^*(x)=\ovl{f(x\inv)}. \label{definv}
\end{equation}
We here use the conventions in \cite{Ren}; many authors include the modular homomorphism
$\Dl:G_1\raw \R_+$ in (\ref{definv}), defined by $\Dl(x)=d\nu(x)/d\nu\inv(x)$. 
We write $L^2(G)$ for $L^2(G_1,\nu)$. 
For $\ps\in L^2(G)$ the formulae
\begin{eqnarray}
\pi_L(f)\ps & = & (\Dl^{-1/2}f)*\ps; \\
\pi_R(f)\ps & = & \ps *f
\end{eqnarray}
define the left and right regular \rep s of $II(G_1)$; one then has
$W^*(G)=\pi_L(II(G_1))''$, which is in standard from with respect to
$J:  L^2(G)\raw L^2(G)$ defined by 
\begeq
J\ps(x)=\Dl(x)^{-1/2}\ps^*(x).
\end{equation}
One then has $JW^*(G)J=W(G)'=\pi_R(II(G_1))''$. 

We have now defined the alleged functor $G\mapsto W^*(G)$ on objects. To define it on
arrows, let $H$ be a second measured groupoid $H$ (with Haar system $\lm$),
and let
$\Ph:G\raw H$ be a measured functor (cf.\ Definition \ref{defmf}).
Define a \Hs\ 
\begeq
L^2(\Ph)=L^2\left(G_0\times^{\Ph_0,t}_{H_0}H_1, \int_{G_0}d\til{\nu}(u)\, \lm^{\Ph_0(u)}\right).
\end{equation}
Compare (\ref{mg1}).  Also, define $\pi_{\lm}: II(G_1)\raw \B(L^2(\Ph))$ 
and  $\pi_{\rh}: II(H_1)\raw \B(L^2(\Ph))$ by
\begin{eqnarray}
\pi_{\lm}(f)\phv(u,h) & = & \int_{G_1} d\nu^u(y)\, \Dl(y)^{-1/2} f(y)\phv(s(y),\Ph_1(y\inv)h);   \\
\pi_{\rh}(g)\phv(u,h) & = & \int_{H_1} d\lm^{s(h)}(l)\, g(l\inv) \phv(u,hl).
\end{eqnarray}
These expressions extend to $f\in W^*(G)$ and $g\in W^*(H)$ by continuity, and it is
easily seen that one thus defines a correspondence
$W^*(G)\rac L^2(\Ph)\lac W^*(H)$.
\begin{theorem}\label{T1}
The map $W^*:\MG\raw\Wa$, defined on objects by $W^*_0(G)=W^*(G)$ as above, and
on arrows (i.e., natural isomorphism classes of measured functors $\Ph:G\raw H$) by 
$$
W^*_1([\Ph])=[W^*(G)\rac L^2(\Ph)\lac W^*(H)],
$$ is a functor.
\end{theorem}

\textbf{Proof}. For $H=G$ and $\Ph=\id$ one easily sees that $L^2(\id)\cong L^2(G)$,
$\pi_{\lm}\cong \pi_L$, and $\pi_{\rh}\cong \pi_R$ (the $\cong$ here standing for
unitary equivalence). Hence one obtains the standard form
$$
W^*_1(\id)=[ W^*(G)\rac L^2(G)\lac W^*(G)].
$$
Since the unit arrows in \Wa\ are precisely the standard forms, this shows that
$W^*$ maps units into units. 

We now need to show that, for a third measured groupoid $K$ and a measured functor
$\Ps:H\raw K$,  one has
\begin{eqnarray}
 W^*(G)\rac L^2(\Ph)\boxtimes_{W^*(H)} L^2(\Ps) \lac W^*(K) & &\\
\cong W^*(G)\rac L^2(\Ps\circ\Ph)\lac W^*(K). & & \label{fuw}
\end{eqnarray}

Since $W^*(H)\rac L^2(H)$ is in standard form, one can easily compute
the relative tensor product by applying the general prescriptions in
\cite{Sau} to the case at hand.  We use the notation in \cite{Sau} and \cite{Hah2}.
Thus $\A_I\subset L^2(H)$ is the left Hilbert algebra associated to 
the above standard form. This defines a normal semi-finite
faithful weight $\lm$ on $W^*(H)$ by $\lm(f^* *f)=\| f\|^2_{L^2(H)}$
for $f\in\A_I$, and $\lm(f^* *f)=\infty$ otherwise. The space of $\lm$-bounded
vectors in $L^2(\Ps)$ is called $D(L^2(\Ps),\lm)$.  One defines a sesquilinear
form on $L^2(\Ph)\ot D(L^2(\Ps),\lm)$ (algebraic tensor product over $\C$)
by sesquilinear extension of
\begeq
(\phv_1\ot\ps_1, \phv_2\ot\ps_2)_0=
(\phv_1, \pi_{\rh}(\la\ps_1,\ps_2\ra_{\lm})\phv_2)_{L^2(\Ph)}, \label{def0}
\end{equation}
where $\la\ps_1,\ps_2\ra_{\lm}\in W^*(H)$ in fact lies in $\A_I$, and may be determined
by its property
\begeq
(f, \la\ps_1,\ps_2\ra_{\lm})_{L^2(H)}=(\ps_1,\pi_{\lm}(Jf)\ps_2)_{L^2(\Ps)}, \label{saup}
\end{equation}
where $f\in\A_I$ is arbitrary. The form $(\, ,\, )_0$ is positive semidefinite, and the completion of
the quotient of $L^2(\Ph)\ot D(L^2(\Ps),\lm)$ by the null space of $(\, ,\, )_0$ in the induced
norm is the \Hs\ $L^2(\Ph)\boxtimes_{W^*(H)} L^2(\Ps) $. The actions
of $W^*(G)$ and $W^*(K)$ on $L^2(\Ph)$ and $D(L^2(\Ps),\lm)\subset L^2(\Ps)$ 
(which is stable under
$W^*(K)$), respectively,  induce  actions on 
$L^2(\Ph)\boxtimes_{W^*(H)} L^2(\Ps) $, defining this \Hs\  as a $W^*(G)$-$W^*(K)$
correspondence.

Denoting the Haar system on $K$ by $\rh$, from (\ref{saup}) one easily finds
\begeq
\la\ps_1,\ps_2\ra_{\lm}(h)=\int_{K_1} d\rh^{\Ps_0(s(h))}(k)\, \ovl{\ps_1(s(h),k)}
\ps_2(t(h),\Ps_1(h)k), \label{predef0}
\end{equation}
from which the form (\ref{def0}) may explicitly be computed. 
Now define $$\til{U}: L^2(\Ph)\ot D(L^2(\Ps),\lm)\raw L^2(\Ps\circ\Ph)$$ by linear extension of
\begeq
\til{U}(\phv\ot\ps):(u,k)\mapsto \int_{H_1} d\lm^{\Ph_0(u)}(h)\, \phv(u,h)\ps(s(h),\Ps_1(h\inv)k).
\end{equation}

Using (\ref{predef0}) and (\ref{def0}), one finds that
\begeq
(\til{U}(\phv_1 \ot\ps_1),\til{U}(\phv_2\ot\ps_2))_{L^2(\Ps\circ\Ph)}=
(\phv_1\ot\ps_1, \phv_2\ot\ps_2)_0.
\end{equation}
Hence $\til{U}$ descends to an isometric  map $U:  L^2(\Ph)\boxtimes_{W^*(H)} L^2(\Ps)
\raw L^2(\Ps\circ\Ph)$. Using the fact that the underlying measure spaces are analytic,
 it is easily shown that the range of $\til{U}$ is dense, so that $U$ 
 is unitary.
A simple computation finally shows that $U$ intertwines the 
pertinent actions of $W^*(G)$ and $W^*(K)$. This proves (\ref{fuw}).
\enp

Since Morita equivalence for measured groupoids is isomorphism
in \MG, and Morita equivalence of \vna s is isomorphism in \Wa,
it follows that the map $G\mapsto W^*(G)$ preserves Morita
equivalence.
\section{Functoriality of $G\mapsto C^*(G)$}\eqnul
\subsection{The category \LG\ of Lie groupoids and principal bibundles}
 Lie groupoids \cite{Mac} play a central role in differential geometry,
once one starts looking for them. This applies, in particular, to foliation
theory \cite{Con1,Con}. In addition, many physical systems can be modeled
by Lie groupoids \cite{NPL3}.
\begin{definition}
A Lie groupoid is a groupoid for which $G_1$
and $G_0$ are manifolds, $s$ and $t$ are surjective submersions, and
$m$ and $I$ are smooth. 
\end{definition}
 It follows that object inclusion is an immersion, that
$I$ is a diffeomorphism, that $G_2$ is a closed submanifold of $G_1\x
G_1$, and that for each $q\in G_0$ the fibers $s\inv(q)$ and
$t\inv(q)$ are submanifolds of $G_1$. In this paper we include
Hausdorffness in the definition of a manifold for simplicity, though
the total space $G_1$ of the holonomy groupoid of a foliation usually fails to satisfy
this condition. With more technical machinery, our results should extend to that case also.

The category \LG, and the key concept of a principal bibundle occurring in its
definition, arose in the work of Moerdijk
\cite{Moe}, originally in the context of topos theory. Similar structures independently
emerged in foliation theory \cite{Con1,Hae,HS}. The connection between these two  points
of entry was made by  Mr\v{c}un \cite{Mrc1,Mrc2}, from which the following
definitions are taken. For the basic underlying notion of a
Lie groupoid action cf.\ \cite{Mac}. 
\begin{definition}\label{alldefs}
A $G$-$H$ bibundle is a manifold $M$ equipped with smooth maps
$M\stackrel{\ta}{\raw} G_0$ and $M\stackrel{\sg}{\raw} H_0$, 
a left $G$-action  $(x,m)\mapsto xm$ from $G
\times^{s,\ta}_{G_0}M$ to $M$, and a right $H$ action 
$(m,h)\mapsto mh$ from $M\times^{t,\ta}_{H_0} H$ to $M$, such that
$\ta(mh)=\ta(m)$, $\sg(xm)=\sg(m)$, and $(xm)h=x(mh)$ for all $(m,h)\in
M\times H$ and $(x,m)\in G\times M$. We write $G\rac M\lac H$.

Such a bibundle is called left principal when $\sg$ is a surjective submersion,
the $G$ action is free (in that $xm=m$ iff $x\in G_0$) and
transitive along the fibers of $\sg$. Equivalently, 
the map from $G_1 \times_{G_0}^{s,\ta}M\raw M\times_{H_0}
M$ given by $(x,m)\mapsto (xm,m)$ is a diffeomorphism.

A $G$-$H$ bibundle $M$ is called regular when it is left principal
and the right $H$ action is proper (in that the map $(m,h)\mapsto
(m,mh)$ from $M\times_{H_0}H$ to $M\x M$ is proper).

Two  $G$-$H$ bibundles $M,N$ are called isomorphic
if there is a diffeomorphism  $M\raw N$
that intertwines the maps $M\raw G_0$, $M\raw H_0$ with the maps
$N\raw G_0$, $N\raw H_0$, and in addition intertwines the $G$ and $H$
actions (the latter condition is well defined because of the former).
\end{definition}

Note that the $G$ action in a left principal $G$-$H$ bibundle is automatically
proper.
In the topos literature a left principal bibundle is seen as a generalized
map from $H$ to $G$, whereas in the foliation literature it is regarded
as the graph of a map between the leaf spaces of the foliations defining
$G$ and $H$.

Now suppose one has left principal bibundles $G\rac M\lac H$ and $H\rac N\lac K$.
The fiber product $M\times_{H_0} N$ 
carries a right $H$ action, given by
$h:(m,n)\mapsto (mh,h\inv n)$ (defined as appropriate). We denote the
 the orbit space by
\begeq
M\otg_H N=(M\times_H N)/H. \label{btp} 
\end{equation}
This is a manifold, and, indeed, a $G$-$K$ bibundle under the obvious maps.
The ``tensor product'' $\otg$ is well defined on isomorphism classes.
The canonical $G$-$G$ bibundle $G$, defined by putting
$M=H=G$, $\ta=t$, and $\sg=s$ in the above definitions, with left and right
actions given by multiplication in the groupoid,
is a left and a right unit for the
bibundle tensor product (\ref{btp}), up to isomorphism.
\begin{definition}\label{defLG}
The category \LG\ has Lie groupoids
as objects and isomorphism classes of regular (i.e., left principal
and right proper) bibundles as
arrows. The arrows  are composed by (\ref{btp}), descending to isomorphism
classes. The units $1_G$ in \Gr\ are the
isomorphism classes $[G\rac G\lac G]$ of the canonical bibundles.
\end{definition}

A number of definitions of Morita equivalence of Lie groupoids have appeared in the literature
\cite{Hae,MRW,Moe,X1,Mrc1,Mrc2}; 
it can be shown that these are all equivalent, and that
two Lie groupoids are Morita equivalent iff they are isomorphic objects in \LG\ \cite{Mrc1,NPLO}.
\subsection{The category \Ca\ of \ca s and Hilbert bibundles}
The definition of \Ca\ is based on the concept of an 
 $\A$-$\B$ Hilbert bimodule, which  is what Rieffel \cite{Rie2}  called
an Hermitian $\B$-rigged $\A$-module, with strict continuity of the $\A$ action
added. Thus an  $\A$-$\B$ Hilbert bimodule
is  a Hilbert $C^*$ module $\CE$ over $\B$, along with a nondegenerate
$\mbox{}^*$-homomorphism of $\A$ into $\CL_{\B}(\CE)$. We write
$\A\rac\CE\rlh\B$. 
Two $\A$-$\B$ Hilbert
bimodules $\CE,\CF$ are called isomorphic when there is a unitary
$U\in\CL_{\B}(\CE,\CF)$; cf.\ \cite{Lance}, p.\ 24.

The canonical bimodule $1_{\B}$ over a \ca\ $\B$ is defined by
$\langle A,B\ra_{\B} =A^*B$, and the left and right actions are given
by left and right multiplication, respectively. 
Rieffel's interior tensor product \cite{Rie2,Lance} maps an
$\A$-$\B$ Hilbert bimodule $\CE$ and a $\B$-$\GC$ Hilbert bimodule
$\CF$ into an $\A$-$\GC$ Hilbert bimodule
$\CE\hat{\ot}_{\B}\CF$. This operation
is well defined on unitary isomorphism classes, and $1_{\B}$ acts as
a two-sided unit for $\hat{\ot}_{\B}$, up to isomorphism.
\begin{definition}\label{defCa}
The category \Ca\ has $C^*$-algebras as objects, and isomorphism
class\-es of Hilbert bimodules as arrows. The arrows are composed by
Rieffel's  interior tensor product, for which the canonical Hilbert
bimodules $1_{\A}$ are   units.
\end{definition}

This category was introduced independently in \cite{Sch},
and, in the guise of a bicategory (where the arrows are Hilbert
bimodules rather than isomorphism classes thereof), in
\cite{NPLDR}. It was shown in \cite{Sch} that two \ca s
are Morita equivalent as defined by Rieffel
\cite{Rie2} iff they are
isomorphic as objects in \Ca; also see \cite{NPLO} for a detailed proof.
The nondegeneracy condition in the definition of the arrows in \Ca\ is essential
for this result.

It should be noted that Thm.\ 2.2 in \cite{BDH} implies that the category
\Wa\ of Definition \ref{defWa} is isomorphic to the subcategory
of \Ca\ consisting of \vna s as objects and normal selfdual Hilbert bimodules as arrows.
\subsection{The map $G\mapsto W^*(G)$ as a functor}
We will now prove that the map $G\mapsto C^*(G)$ mentioned in the Introduction 
may be extended
so as to associate Hilbert bimodules to regular bibundles, thus defining
a functor from \LG\ to \Ca. Although it should be possible to use the geometric
definition of $C^*(G)$ in terms of half-densities \cite{Con}, as in our
previous direct proof that $G\mapsto C^*(G)$ preserves Morita equivalence
\cite{NPLvE}, we find it
much easier to regard a Lie groupoid as a locally compact groupoid
with smooth Haar system (cf.\ the Introduction).

Specifically, a Lie groupoid $G$ has a
left Haar system $\{\nu^q\}_{q\in G_0}$ such that $\nu^q$ 
 is supported on $t\inv(q)$ and is equivalent to
Lebesgue measure in each coordinate chart (recall that $t\inv(q)$ 
is a submanifold of $G_1$). Furthermore, for each $f\in\cci(G_1)$ the
function $q\mapsto \int d\nu^q(x)\, f(x)$ on $G_0$ is smooth. 
This endows $\cci(G)$ with the structure of a \sta\ under the operations
(\ref{defconv}) and (\ref{definv}).
The groupoid \ca\ $C^*(G)$ is a suitable completion of the \sta\ $\cci(G)$;
see \cite{Ren} for the analogous case of $C_c(G)$, or \cite{Con,NPL3} for
the smooth case.

We have now defined the map $G\mapsto C^*(G)$ on objects. To define it
on arrows,
let  $G\rac M\lac H$ be a regular bibundle (cf.\ Definition \ref{alldefs}
 for the notation that will be used throughout this chapter).
A key fact is that a Haar system on $G$ defines a family of measures
$\{\mu^r\}_{r\in H_0}$ on $M$, where
$\mu^r$ is supported on $\sg\inv(r)$, on which it is  equivalent to
Lebesgue measure in each coordinate chart. Moreover, for each
$f\in\cci(M)$ the
function $r\mapsto \int d\mu^r(m)\, f(m)$ on $H_0$ is smooth,
and the family is
$H$-equivariant (in the sense of \cite{Ren87}) with respect to
$\sg$, the given $H$ action on $M$, and the natural right $H$ action on $H_0$.
This means that for each $f\in\cci(M)$ one has
\begeq
\int d\mu^{t(h)}(m)\, f(mh)=\int d\mu^{s(h)}(m)\, f(m). \label{nuHeq}
\end{equation}
Namely, for fixed $r\in H_0$ this system is defined by choosing
$m_0\in\sg\inv(r)$, and putting
\begeq
\int d\mu^r(m)\, f(m)=\int d\nu^{\ta(m_0)}(x)f(x\inv m_0). \label{defnu}
\end{equation}
Using (\ref{LHM}), one verifies that this is independent of the choice
of $m_0$ (despite the fact that $\ta(m_0)$ is not constant on
$\sg\inv(r)$). This definition is evidently possible because in a
regular bibundle the $G$ action is principal over $\sg$.

The following lemma is similar to Thm.\ 2.8 in \cite{MRW}, and also
appeared in \cite{SO} for the locally compact case
(this paper was drawn to our attention after the
circulation of an earlier draft of this paper as an e-print); our
assumptions are weaker, since we do not have an equivalence bibundle
but merely a regular one.  However, what is really used in \cite{MRW}
is precisely our regularity properties.
\begin{lemma}\label{MRWlem}
Let $G\rac M\lac H$ be a regular bibundle. The formulae
\begin{eqnarray}
\la\phv,\ps\ra & : & h\mapsto \int d\mu^{t(h)}\, \ovl{\phv(m)}\ps(mh); 
\label{F1}\\
f\cdot\phv & : & m\mapsto \int d\nu^{\ta(m)}\, (x)f(x)\phv(x\inv
m);\label{F2}\\
\phv\cdot g & : & m\mapsto \int d\lm^{\sg(m)}\, g(h\inv)\phv(mh),\label{F3}
\end{eqnarray}
where $\phv,\ps\in\cci(M)$, $f\in\cci(G)$, and $g\in\cci(H)$,
define functions in $\cci(H)$, $\cci(M)$, and $\cci(M)$, respectively.
This equips $\cci(M)$ with the structure of a pre Hilbert $C^*$-module
over $\cci(H)$ (seen as a dense subalgebra of $C^*(H)$), on which
$\cci(G)$ (seen as a dense subalgebra of $C^*(G)$) acts nondegenerately
by adjointable operators.  This structure may be completed to a 
$C^*(G)$-$C^*(H)$ Hilbert bimodule, which we call $\CE(M)$.
\end{lemma}

\textbf{Proof} 
It should now be obvious why the right $H$ action on a regular bibundle
has to be proper, since this guarantees $\cci(H)$-valuedness of the inner
product (otherwise, one could land in $\cin(H)$).

The necessary algebraic properties may be checked by elementary
computations. The property $\la\phv,\ps\ra^*=\la\ps,\phv\ra$ follows
from (\ref{nuHeq}), the property $\la\phv,\ps\cdot g\ra=\la\phv,\ps\ra*g$
is an identity, the properties $\la\phv,f\cdot\ps\ra=\la f^*\cdot\phv,\ps\ra$
and $(f_1*f_2)\cdot \phv=f_1\cdot (f_2\cdot\phv)$
require (\ref{defnu}) and (\ref{LHM}), and finally $\phv\cdot(g_1*g_2)=
(\phv\cdot g_1)\cdot g_2$ follows from (\ref{LHM}) for $\lm$.

The proof of positivity of $\la\, ,\,\ra$ is the same as in \cite{MRW};
it follows from Prop.\ 2.10 in \cite{MRW} and the argument of P. Green
(Remark following Lemma 2 in \cite{Gre}). This also proves the nondegeneracy
of the action of $\cci(G)$ (and hence of the ensuing action of $C^*(G)$).

We cannot use the entire argument in \cite{MRW} to the effect 
that everything
can be completed, since in \cite{MRW} one has a $\cci(G)$-valued inner
product as well. However, it is quite trivial to proceed, since by the above
results $\cci(M)$ is a pre Hilbert $C^*$-module
over $\cci(H)$, which can be completed to a Hilbert $C^*$-module
$\CE(M)$ over $C^*(H)$ in the standard way (cf.\ Ch.\ 1 in \cite{Lance}
or Cor.\ IV.2.1.4 in \cite{NPL3}). One then copies the proof in \cite{MRW}
of the property
\begeq
\la f\cdot\phv,f\cdot\phv\ra \leq \| f\|^2 \la\phv,\phv\ra, \label{doublestar}
\end{equation}
where the norm is in $C^*(G)$,  to complete
the argument.
\enp

\begin{theorem}\label{Thm1}
The map  $C^*:\LG\mapsto\Ca$, defined on objects  by $C^*_0(G)=C^*(G)$, and
on arrows by $$C^*_1([G\rac M\lac H])=[C^*(G)\rac \CE(M)\rlh C^*(H)],$$
is a functor.
\end{theorem}

\textbf{Proof} 
We begin with the unit arrows. We claim that
the construction in Lemma \ref{MRWlem} maps the canonical bibundle
$G\rac G\lac G$ into the canonical Hilbert bimodule $C^*(G)\rac
C^*(G)\rlh C^*(G)$.
It is easy to check from
(\ref{F1}) - (\ref{F3}) that $\la\phv,\ps\ra=\phv^* *\ps$,
$f\cdot\phv=f*\phv$, and $\phv\cdot g=\phv*g$. These properties
pass to the completions by continuity. Hence $C^*$ preserves units.

Now let $H\rac N\lac K$ be a second regular bibundle, so that one may form
the bibundle tensor product $M\otg_H N$ (cf. (\ref{btp}))
and its associated $C^*(G)$-$C^*(K)$ Hilbert bimodule $\CE(M\otg_H N)$.
To compare this with the $C^*(G)$-$C^*(K)$ Hilbert bimodule 
$\CE(M)\otq_{C^*(H)}\CE(N)$, we 
define a map $\til{U}:\cci(M)\ot_{\C}\cci(N)\raw\cci(M\otg_H N)$
by
\begeq
\til{U}(\phv\ot_{\C}\ps): [m,n]_H\mapsto \int d\lm^{\sg(m)}(h)\,
 \phv(mh)\ps(h\inv n).
\end{equation}

Note that the right-hand side is well defined on $[m,n]_H$ rather than
$(m,n)$ because of the invariance property (\ref{LHM}) for $H$.  This
map was introduced by Mr\v{c}un \cite{Mrc1} for smooth \'{e}tale
groupoids; we have merely replaced the counting measure by a general
Haar system.

We now show that
the map $\til{U}$ leaves the kernel of the canonical projection 
$$
\cci(M)\ot_{\C}\cci(N)\raw\CE(M)\otq_{C^*(H)}\CE(N)$$
stable, that $\til{U}$ has dense range, and that
accordingly  the corresponding quotient map $U$, extended by continuity,
defines an isomorphism  
\begin{equation}
\CE(M)\otq_{C^*(H)}\CE(N)\simeq\CE(M\otg_H N) \label{forgeq}
\end{equation}
 as $C^*(G)$-$C^*(K)$ Hilbert bimodules.

A lengthy but straightforward computation shows that
$$
\la \til{U}(\phv_1\ot_{\C}\ps_1, \til{U}(\phv_2\ot_{\C}\ps_2)\ra_{C^*(K)}^{\CE(M\otg_H N)},
$$
is equal to
$$
\la\ps_1,\la\phv_1,\phv_2\ra^{\CE(M)}_{C^*(H)}\cdot\ps_2\ra^{\CE(N)}_{C^*(K)},
$$
which by definition is equal to
$$
\la \phv_1\ot_{C^*(H)}\ps_1,\phv_2\ot_{C^*(H)}\ps_2\ra_{C^*(K)}^{\CE(M)\otq_{C^*(H)}\CE(N)}.
$$
Here $\phv\ot_{C^*(H)}\ps$ is the image of $\phv\ot_{\C}\ps$ in $\CE(M)\otq_{C^*(H)}\CE(N)$.
In view of the definitions of the various Hilbert $C^*$-modules over $C^*(K)$ involved,
this computation implies that $\til{U}$ quotients and extends to an isometry
$U$ from $\CE(M)\otq_{C^*(H)}\CE(N)$ to $\CE(M\otg_H N)$. 

Moreover, using the fact that $M$ and $N$ are manifolds,
it is easily seen that $\til{U}$ has a dense
range in 
$\cci(M\otg_H N)$ with respect to the inductive limit
topology, so that it certainly has a dense range for the topology induced 
on  $\cci(M\otg_H N)$ by
the norm on $\CE(M\otg_H N)$ as a Hilbert $C^*$-module over $C^*(K)$ (since the latter
topology is finer than the former). Since  $\cci(M\otg_H N)$ is itself dense in
$\CE(M\otg_H N)$ in the latter topology, it follows that 
$\til{U}$ has dense range when seen as a map taking values in $\CE(M\otg_H N)$.
Hence $U$ is an isometric isomorphism between 
$\CE(M)\otq_{C^*(H)}\CE(N)$ and $\CE(M\otg_H N)$
as Banach spaces.

Another elementary computation shows that 
\begeq
\til{U}(\phv\ot_{\C}(\ps\cdot g))=\til{U}(\phv\ot_{\C}\ps)\cdot g \label{Urechts}
\end{equation}
for $\phv\in\cci(M)$, $\ps\in\cci(N)$, and $g\in\cci(H)$. This implies that
\begeq
U(\phv\ot_{C^*(H)}(\ps\cdot g))=U(\phv\ot_{C^*(H)}\ps)\cdot g \label{Ur2}
\end{equation}
for all $\phv\in\CE(M)$, $\ps\in\CE(N)$, and $g\in C^*(H)$.
The reason for this is that a continuous $\B_0$-linear map
between two pre Hilbert $C^*$ modules over a dense subalgebra
$\B_0$ of $\B$  extends to a $\B$-linear map between the completions;
this easily follows from the bound $\|\ps B\|\leq \| B\|\,\|\ps\|$.

We conclude that $U$ is a $C^*(K)$-linear isometric isomorphism, and hence
by Thm.\ 3.5 in \cite{Lance} it is actually unitary (in particular, it now follows
that $P$ is adjointable). 

Finally, analogously to (\ref{Urechts}) one obtains the equality
\begeq
\til{U}(f\cdot(\phv\ot_{\C}\ps)=f\cdot \til{U}(\phv\ot_{\C}\ps),\label{Ulinks}
\end{equation}
where $f\in\cci(G)$. This time, the passage of this property to
the pertinent completions is achieved through (\ref{doublestar}), which leads to
the bound $\| A\ps\|\leq\| A\|\,\|\ps\|$ for any adjointable operator
on a (pre) Hilbert $C^*$-module. Thus $U$ is $C^*(G)$-linear as well.
This proves (\ref{forgeq}).

Hence $C^*$ preserves composition of arrows, and Theorem \ref{Thm1} follows.
\enp

Since Morita equivalence of Lie groupoids is isomorphism in \LG, and
Morita equivalence of \ca s is isomorphism in \Ca, we recover the
known result that the map $G\mapsto C^*(G)$ preserves Morita equivalence
\cite{MRW,NPLvE}.
\section{Functoriality of $G\mapsto A^*(G)$}\eqnul
The category on which the map $A^*$ is going to be defined is 
as follows.
\begin{definition}\label{defLGb}
The category \LGc\ has s-connected and s-simply connected Lie
groupoids as objects, and isomorphism classes of left principal
bibundles as arrows. The arrows and units are as in Definition
\ref{defLG}.
\end{definition}

In contrast with Definition \ref{defLG}, the class of objects is more restricted;
this will be necessary for our functor to preserve units. On the other hand, the bibundles need
not be right proper.
\subsection{The category \Po\ of Poisson manifolds and dual pairs}
The definition of a suitable category of Poisson manifolds \cite{NPLO} is based on the
theory of symplectic groupoids (cf.\ \cite{CDW,Vai} and refs.\ therein).
The objects in \Po\ are defined as follows.
\begin{definition}\label{defintP}
A Poisson manifold $P$ is called integrable when there
exists an $s$-connected and $s$-simply connected
symplectic groupoid $\Gm(P)$ over $P$.
\end{definition}
This definition has been adapted from \cite{CDW}, where no connectedness
requirements are made. When it exists, such a symplectic groupoid can be shown to be unique
up to isomorphism (see \cite{MM} for Lie groupoids and
\cite{NPLO} for symplectic groupoids). It is easily seen that a Poisson manifold is integrable
iff it is Morita equivalent to itself in the sense of Xu \cite{X1}.

The arrows in \Po\ will be isomorphism classes of certain dual pairs.
Given two Poisson manifolds
$P$ and $Q$,  a dual pair $Q\law S \raw P$ consists of a
symplectic manifold $S$ and Poisson
maps $q:S\raw Q$ and $p:S\raw P^-$, such
that $\{q^* f, p^* g\}=0$ for all $f\in\cin(Q)$ and $g\in\cin(P)$ \cite{W1,K1}.
In a complete dual pair the maps $p$ and $q$ are complete;
a Poisson map $J:S\raw P$ is called complete when, for every
$f\in\cin(P)$ with complete Hamiltonian flow, the Hamiltonian
flow of $J^*f$ on $S$ is complete as well (that is, defined for
all times). Two $Q$-$P$ dual pairs
$Q\stackrel{q_i}{\law}\til{S}_i\stackrel{p_i}{\raw}P$, $i=1,2$,
are isomorphic when there is a symplectomorphism 
$\phv:\til{S}_1\raw \til{S}_2$ for which $q_2\phv=q_1$ and $p_2\phv=p_1$.

Based on results in \cite{CDW,Daz,X2}, it can be shown that
for integrable Poisson manifolds $P$ and $Q$, with associated
s-connected and s-simply connected symplectic groupoids $\Gm(P)$ and
$\Gm(Q)$, there is a natural bijective correspondence between
complete dual pairs $Q\law S\raw P$ and 
symplectic bibundles $\Gm(Q)\rac S \lac\Gm(P)$.
In particular, the canonical symplectic bibundle
associated to the dual pair $P\stackrel{t}{\law}\Gm(P)\stackrel{s}{\raw}P$ is
$\Gm(P)\rac\Gm(P)\lac\Gm(P)$.
Accordingly, we say that a dual pair is regular when it is complete and when the
associated symplectic bibundle is left principal
(it is not necessary to impose properness of the right $\Gm(P)$ action).

Let $R$ be a third integrable Poisson manifold, with associated
s-connected and s-simply connected symplectic groupoid $\Gm(R)$,
and let $Q\law S_1\raw P$ and 
$P\law S_2\raw R$ be regular dual pairs. The embedding $S_1\x_P S_2\subset S_1\x S_2$
is coisotropic \cite{NPL3}; we denote the corresponding symplectic quotient by
$S_1\otc_{P}S_2$. This is the middle space of a regular dual pair
$P\law S_1\otc_{P} S_2\raw R$, which we regard as the tensor product of
the given dual pairs. An alternative way of defining this tensor product
is to  construct the groupoid tensor product $\Gm(Q)\rac  S_1\circledast_{\Gm(P)} S_2\lac\Gm(R)$
of the associated symplectic bibundles \cite{X1}. 
Thus we have
\begeq
S_1\otc_{P} S_2=S_1\circledast_{\Gm(P)} S_2 \label{rhsxu}
\end{equation}
as symplectic manifolds, as $\Gm(Q)$-$\Gm(R)$ symplectic bibundles, and as
 $Q$-$R$ dual pairs.
In any case, this tensor
product is associative up to isomorphism, and the dual pair
$P\stackrel{t}{\law}\Gm(P)\stackrel{s}{\raw}P$ is a two-sided unit for $\otc_P$,
up to isomorphism \cite{NPLO}.
\begin{definition}\label{defPo}
The category \Po\ has integrable Poisson manifolds as objects, and isomorphism
classes of regular dual pairs as arrows. The arrows are composed by
the tensor product $\otc$, for which the dual pairs
$P\stackrel{t}{\law}\Gm(P)\stackrel{s}{\raw}P$ are units.
\end{definition}

The original reason for the introduction of this category was not so much
the subsequent functoriality theorem, but rather the fact that 
two  Poisson manifolds are  Morita equivalent in the sense of Xu \cite{X2}
iff they are isomorphic objects in \Po\ \cite{NPLO}. Moreover, we 
now have a classical analogue of the categories \Wa\ and \Ca.
\subsection{The map $G\mapsto A^*(G)$ as a functor}
A Lie groupoid $G$ defines an associated ``infinitesimal'' object, its
Lie algebroid $A(G)$ \cite{Pra}; see \cite{Mac,CDW,NPL3} for
reviews. The main point is that $A(G)$ is a vector bundle over $G_0$,
endowed with an ``anchor map'' $\al:A(G)\raw T(G_0)$ and a Lie algebra
structure on its space of sections $\cin(G_0,A(G))$ that is compatible
with the anchor map in a certain way. 

It is of central importance to
us that the dual vector bundle $A^*(G)$ is a Poisson manifold in a
canonical way \cite{CDW,Cou,DS},
which generalizes the well-known Lie--Poisson structure on the dual of
a Lie algebra. We look at the passage $G\mapsto A^*(G)$ as a classical
analogue of the map $G\mapsto C^*(G)$. 

Another important construction is that of the cotangent bundle $T^*(G)$
of $G$. This is not merely
a symplectic space (equipped, in our conventions \cite{NPL3,NPLvE}, 
with minus the usual symplectic form on a cotangent bundle, so that we 
write $T^*(G)^-$ when this aspect is relevant), but a symplectic groupoid
with $T^*(G)_1=T^*(G_1)$ 
over $T^*(G)_0=A^*(G)$ \cite{CDW} (also see \cite{Vai} for a review).
For simplicity we will write $T^*(G)$ for $T^*(G_1)$, and
 denote the source and target projections of $T^*(G)$ by
$\til{s}$ and $\til{t}$, respectively.
\begin{lemma}\label{MMcor}
The Poisson manifold $A^*(G)$ associated to a Lie groupoid $G$ is
integrable (cf.\ Definition \ref{defintP}).
\end{lemma}

\textbf{Proof}  We use
 Prop.\ 3.3 in \cite{MM} to infer the existence of an s-connected and
 s-simply connected Lie groupoid $\til{G}$ for which
 $A(G)=A(\til{G})$.  Since the Poisson structure on $A^*(G)$ is
 entirely determined by the Lie algebroid structure of $A(G)$, one has
 $A^*(G)=A^*(\til{G})$.  It may be checked from its definition that
 $T^*(G)$ is s-connected and s-simply connected iff $G$ is, so that
 $T^*(\til{G})$ is a s-connected and s-simply symplectic groupoid over
 $A^*(G)$.
\enp

In view of this lemma, we will henceforth assume that all Lie groupoids
are s-connected and s-simply connected (dropping the tilde).
Thus we have defined the map $A^*: \LGc\raw\Po$ on objects.

In order to define this map on arrows,
we recall a number of results from \cite{NPLvE}, which we here 
combine into a lemma. 
\begin{lemma}\label{MRWcl}
Any bibundle $G\rac M\lac H$ (cf.\ Definition \ref{alldefs})
defines a symplectic bimodule
\begeq A^*(G)\stackrel{J^G_L}{\longleftarrow} T^*(M)^-
\stackrel{J^H_R}{\longrightarrow}A^*(H),
\end{equation}
with associated symplectic bibundle 
\begeq
T^*(G)^-\rac T^*(M)^-\lac T^*(H)^-.
\end{equation}
\end{lemma}

The explicit form of the ``momentum map'' 
$J_R^H$ is
\begeq
\left\langle J^H_R(\theta_m),
\frac{dh(\lm)}{d\lm}_{|\lm=0}\right\rangle_{\sg(m)}
 =
\left\langle\theta_m,\frac{d m h(\lm)}{d\lm}_{|\lm=0}
\right\rangle_m,\label{JRE}
\end{equation}
where $\theta_m\in T^*_m(M)$, $\sg(m)=h(0)$, and $h(\lm)\in t\inv(\sg(m))$,
so that $\dot{h}(0)$ lies in $A_{\sg(m)}(H)$ and 
$J^H_R(\theta_m)\in A^*_{\sg(m)}(H)$.

The associated right action of $T^*(H)$ on $T^*(M)$ is given by
\begin{equation}
\left\langle \theta_m\cdot (\al_h)\inv,
 \frac{dm(\lm)}{d\lm}_{|\lm=0}\right\rangle_{mh\inv}=
\left\langle\theta_m, \frac{dm(\lm)\til{h}(\lm)}{d\lm}_{|\lm=0}\right\rangle_m
-\left\langle\al_h,\frac{d\til{h}(\lm)}{d\lm}_{|\lm=0}\right\rangle_h,
\label{defTGA}
\end{equation}
where $m(0)=mh\inv$, and $\til{h}(\cdot)$ is a curve in $H$
satisfying $\til{h}(0)=h$ and $\sg(m(\lm))=t(\til{h}(\lm))$.
As explained in \cite{NPLvE}, eq.\ (\ref{defTGA})  is independent of the 
choice of $\til{h}$ because of the compatibility condition
$J^H_R(\theta_m)=\til{s}(\al_h)$ under which $\theta_m\cdot (\al_h)\inv$
is defined; cf.\ Definition \ref{alldefs}. Explicitly, this condition
reads $\sg(m)=s(h)$, along with
\begin{equation}
\left\langle\theta_m, \frac{dm\chi(\lm)}{d\lm}_{|\lm=0}\right\rangle_m=
\left\langle\al_h,\frac{dh\chi(\lm)}{d\lm}_{|\lm=0}\right\rangle_h,
\label{isis}
\end{equation}
for all curves $\ch(\cdot)\in t\inv(s(h))$  subject to $\ch(0)=s(h)$.
Note that these formulae for right actions are not given in \cite{NPLvE},
but they may be derived from those for left actions, together with
the formula $\al\inv=-I^*(\al)$ for the inverse in $T^*(G)$
(where $I:G_1\raw G_1$ is the inverse in $G$) \cite{CDW}.

The explicit form of $J_L$ will shortly be needed not 
for $G\rac M\lac H$, but for a second bibundle
$H\rac N\lac K$; hence we state it for the latter.
The momentum map $J_L^H: T^*(N)\raw A^*(H)$, then, is given by
\cite{NPLvE}
\begeq
\left\langle J^H_L(\et_n),\frac{dh(\lm)}{d\lm}_{|\lm=0}\right\rangle_{\rh(n)}
= 
-\left\langle\et_n,\frac{dh(\lm)\inv
n}{d\lm}_{|\lm=0}\right\rangle_h, \label{Jrev2}
\end{equation}
where $\et_n\in T^*_n(N)$, $\rh(n)=h(0)$, and
$h(\lm)\in t\inv(\rh(n))$; recall that $\rh:N\raw H_0$ is the base map
of the $H$ action on $N$. 

The associated left action of $T^*(H)$ on $T^*(N)$ is given by
\begin{equation}
\left\langle \al_h\cdot\et_n,
 \frac{dn(\lm)}{d\lm}_{|\lm=0}\right\rangle_{hn}=
\left\langle\eta_n, 
\frac{d\hat{h}(\lm)\inv n(\lm)}{d\lm}_{|\lm=0}\right\rangle_n
+\left\langle\al_h,\frac{d\hat{h}(\lm)}{d\lm}_{|\lm=0}\right\rangle_h,
\label{defTGA2}
\end{equation}
where $n(0)=hn$, and $\hat{h}(\cdot)$ is a curve in $H$
satisfying $\hat{h}(0)=h$ and $\rh(n(\lm))=t(\hat{h}(\lm))$.
The condition under which $\al_h\cdot\et_n$
is defined is
$J^H_L(\et_n)=\til{s}(\al_h)$, which
reads $\rh(n)=s(h)$, along with
\begin{equation}
-\left\langle\eta_n, \frac{d\chi(\lm)\inv n}{d\lm}_{|\lm=0}\right\rangle_n
=
\left\langle\al_h,\frac{d\chi(\lm)}{d\lm}_{|\lm=0}\right\rangle_h,
\label{isis2}
\end{equation}
for $\ch$  as specified after (\ref{isis}).
This completes the  exposition of Lemma \ref{MRWcl}.
\begin{theorem}\label{Thm2}
The map $A^*:\LGc\raw\Po$, defined on objects by $A^*_0(G)=A^*(G)$ and
on arrows by
$$
A^*_1([G\rac M\lac H])=[A^*(G)\law T^*(M)^-\raw A^*(H)],
$$
is a functor.
\end{theorem}

\textbf{Proof} The object map $A^*_0$ is well defined between the given
categories by Lemma \ref{MMcor}. Turning to the unit arrows,
we note that the construction in Lemma \ref{MRWcl} maps
the canonical bibundle $G\rac G\lac G$ into the 
symplectic bimodule 
$$
A^*(G)\stackrel{\til{t}}{\law} T^*(G)^- \stackrel{\til{s}}{\raw}A^*(G).
$$
 To see this, recall that $\til{s}$ and $\til{t}$ are the source and target
maps of the symplectic groupoid $T^*(G)^-$. The lemma follows because,
as already remarked in \cite{NPLvE}, $\til{s}$ and $\til{t}$
as defined in \cite{CDW} coincide with the momentum mappings
$J^G_R$ and $J^G_L$ defined by Lemma \ref{MRWcl}, applied to the
canonical bibundle. 
 It is here that the assumption of s-connectedness and s-simply
connectedness is essential. 

We now turn to the composition of arrows.
Let $G\rac M\lac H$ and $H\rac N\lac K$ be regular bibundles, with associated
symplectic bimodules $A^*(G)\law T^*(M)^-\raw A^*(H)$ and
$A^*(H)\law T^*(N)^-\raw A^*(K)$, respectively (cf.\ Lemma \ref{MRWcl}).
We will prove that the tensor product
\begeq
A^*(G)\law T^*(M)^-\otc_{A^*(H)} T^*(N)^-\raw A^*(K) \label{isolhs}
\end{equation}
of these symplectic bimodules is isomorphic to the symplectic bimodule
\begeq
A^*(G)\law T^*(M\otg_H N)^-\raw A^*(K)\label{isorhs}
\end{equation}
associated with the bibundle $G\rac M\otg_H N\lac K$.

We  omit all suffixes ``-'' (as in $S^-$), unless
strictly necessary. 
By  (\ref{rhsxu}) and (\ref{btp}) we have
\begeq
T^*(M)\otc_{A^*(H)} T^*(N)= (T^*(M)*_{A^*(H)} T^*(N))/T^*(H) \label{4lc}
\end{equation}
as $A^*(G)$-$A^*(K)$ symplectic bimodules. By (\ref{btp}), one has
\begeq
T^*(M\otg_H N)=T^*((M*_{H_0}N)/H),
\end{equation}
so we start by proving that
\begeq
(T^*(M)*_{A^*(H)} T^*(N))/T^*(H)\simeq T^*((M*_{H_0}N)/H) \label{siso}
\end{equation}
as symplectic manifolds. 
To do so, we first show that
\begeq
T^*((M*_{H_0}N)/H)\simeq (T^*(M)*_{A^*(H)} T^*(N))/\sim \label{sim1}
\end{equation}
as manifolds, where $\sim$ is an equivalence relation defined as 
follows. For 
 $(\theta_m,\eta_n)\in T^*(M)*_{A^*(H)} T^*(N)$ (i.e.,
$\sg(m)=\rh(n)$ and 
$J^H_R(\theta_m)=J^H_L(\eta_n)$), $h\in s\inv(\sg(m))$, and
$(\theta_{mh\inv}',\eta_{hn}')\in T^*(M)*_{A^*(H)} T^*(N)$, we say that
$(\theta_{mh\inv}',\eta_{hn}')\sim (\theta_m,\eta_n)$ iff for each 
pair of vectors $\dot{m}(0)\in T_{mh\inv}(M)$ and
$\dot{n}(0)\in T_{hn}(N)$ such that 
\begeq
\sg_*(\dot{m}(0))=\rh_*(\dot{m}(0)), \label{starstar}
\end{equation}
there exists a curve $h(\cdot)$ in $H$ with $h(0)=h$ and $t(h(\lm))=
\sg(m(\lm))=\rh(n(\lm))$ (the latter equality may be imposed 
for convenience because of (\ref{starstar})), such that
\begin{eqnarray}
\left\langle \theta_{mh\inv}',
 \frac{dm(\lm)}{d\lm}_{|\lm=0}\right\rangle_{mh\inv}
& + &
 \left\langle\eta_{hn}' ,
 \frac{dn(\lm)}{d\lm}_{|\lm=0}\right\rangle_{mh\inv}= \nn \\
\left\langle \theta_m, \frac{dm(\lm)h(\lm)}{d\lm}_{|\lm=0}\right\rangle_m
& + &
\left\langle \eta_n, \frac{dh\inv (\lm)n(\lm)}{d\lm}_{|\lm=0}\right\rangle_n.
\label{pluspre}
\end{eqnarray}

To prove (\ref{sim1}), note that for any (possibly singular) smooth
foliation $\Ph$ of a manifold $Q$ with smooth leaf space $Q/\Ph$
one has an isomorphism
\begeq
\cin(Q/\Ph, T^*(Q/\Ph))\simeq \cin(Q,T(\Ph)_0^0),\label{fol1}
\end{equation}
where the right-hand side consists of all 1-forms $\om$ on $Q$ that
satisfy $i_{\xi}\om=0$ (forming $T(\Ph)^0\subset T^*(Q)$) and
$i_{\xi}d\om=0$ (defining $T(\Ph)_0$), for all $\xi\in\cin(Q,T(\Ph))$.
This is well known for regular foliations (cf.\ \cite{Mol}), 
and the proof is the same in the singular case (it merely depends on
the smoothness of the leaf space). These conditions may
be rewritten as $i_{\xi}\om=\CL_{\xi}\om=0$ (where $\CL$ is the
Lie derivative), or as $i_{\xi}\om=0$ for all vector fields $\xi$ as above 
and $\phv^*\om=\om$ for all diffeomorphisms $\phv$ of $Q$
that are generated by such $\xi$. The isomorphism (\ref{fol1})
is then given by $\al\leftrightarrow\pi^*\al$, where $\pi:Q\raw Q/\Ph$
is the canonical projection. 
In addition, one has
\begeq
\cin(Q,T(\Ph)^0_0)\simeq  \cin(Q/\Ph, T(\Ph)^0/\sim),\label{fol2}
\end{equation}
where the equivalence relation $\sim$ on $T(\Ph)^0$ is defined by
$\bt'\sim \bt$ iff $\bt'=\phv^*\bt$ for some diffeomorphism $\phv$ as
specified above. The isomorphism (\ref{fol2}) 
associates a section $q\mapsto\bt(q)$ with a section 
$[q]_{\Ph}\mapsto [\bt(q)]_{\sim}$. Hence the ensuing isomorphism
\begeq
\cin(Q/\Ph, T^*(Q/\Ph))\simeq \cin(Q/\Ph, T(\Ph)^0/\sim)\label{fol3}
\end{equation}
is given by $\al\leftrightarrow[\pi^*\al]_{\sim}$. 

We apply this to $Q=M*_{H_0}N$, where $\Ph$ is the foliation by the orbits
of the diagonal $H$ action. The condition of lying in $T(\Ph)^0$
then has $T^*(M)*_{A^*(H)} T^*(N)$ as its solution set, and the
equivalence relation $\sim$ defined for $\Ph$ is precisely the one
imposed by (\ref{pluspre}) and preceding text.
This proves (\ref{sim1}).

Next, we show that the equivalence relation $\sim$ on
$T^*(M)*_{A^*(H)} T^*(N)$ coincides with $\backsim$, defined as
follows.  We say that $(\theta_{mh\inv}',\eta_{hn}')\backsim
(\theta_m,\eta_n)$ iff there exists $\al_h\in T^*_h(H)$ satisfying
\begeq
\til{s}(\al_h)=J^H_R(\theta_m) \label{alhc}
\end{equation}
 (and therefore also
$\til{s}(\al_h)=J^H_L(\eta_n)$), such that for
\textit{each} 
pair of vectors $\dot{m}(0)\in T_{mh\inv}(M)$ and
$\dot{n}(0)\in T_{hn}(N)$ (not necessarily satisfying
(\ref{starstar})), 
there exist curves $\hat{h}(\cdot)$
and $\til{h}(\cdot)$ in $H$ subject to 
  $\hat{h}(0)=\til{h}(0)=h$, $t(\til{h}(\lm))=
\sg(m(\lm))$, $t(\hat{h}(\lm))=\rh(n(\lm))$,  for which one has
\begin{eqnarray}
\left\langle \theta_{mh\inv}',
 \frac{dm(\lm)}{d\lm}_{|\lm=0}\right\rangle_{mh\inv}
& + &
 \left\langle\eta_{hn}' ,
 \frac{dn(\lm)}{d\lm}_{|\lm=0}\right\rangle_{mh\inv}= \nn \\
\left\langle \theta_m, \frac{dm(\lm)\til{h}(\lm)}{d\lm}_{|\lm=0}\right\rangle_m
& + &
\left\langle \eta_n, \frac{d\hat{h}\inv 
(\lm)n(\lm)}{d\lm}_{|\lm=0}\right\rangle_n + \nn \\
\left\langle \al_h, \frac{d\hat{h}(\lm)}{d\lm}_{|\lm=0}\right\rangle_h
& - &
\left\langle \al_h, \frac{d\til{h}(\lm)}{d\lm}_{|\lm=0}\right\rangle_h.
\label{pluspre2}
\end{eqnarray}
We stress that $\hat{h}$ and $\til{h}$ do, and $\al_h$ does
not depend on  the vectors $\dot{m}(0)$ and $\dot{n}(0)$.
The full right-hand side of (\ref{pluspre2}) is independent of
the choice of $\hat{h}$ and $\til{h}$; cf.\ the comment following
(\ref{isis}). 

First, $\backsim$ implies $\sim$ (i.e., $A\backsim B\raw A\sim B$),
for if (\ref{starstar}), and hence $\sg(m(\lm))=\rh(n(\lm))$, holds,
one may choose $h=\til{h}=\hat{h}$, and the final line in (\ref{pluspre2})
drops out, implying (\ref{pluspre}).

To prove the converse, we note that, since the bibundle $G\rac M\lac H$ 
is regular, the map $\sg:M\raw H_0$ is a surjective submersion, so that
$$T_m(M)\simeq T^{\sg}_m(M)\oplus T_{\sg(m)}(H_0).$$
Here $T^{\sg}_m(M)$ is the kernel of $\sg_*:T(M)\raw T(H_0)$ at $m$.
This induces the decomposition
\begeq
T_m(M)\oplus T_n(N)\simeq T_{(m,n)}^{\sg=\rh}(M\x N)\oplus T_{\sg(m)}(H_0),
\label{decTMN}
\end{equation}
where $T_{(m,n)}^{\sg=\rh}(M\x N)$ is the kernel of $\sg_*-\rh_*$
at $(m,n)$. Explicitly, the decomposition of a given
vector according to  (\ref{decTMN}) reads
$$
(\xi_1,\xi_2,\zeta)=(\xi_1,\rh_*(\zeta),\zeta)+(0,\xi_2-\rh_*(\zeta),0),
$$
where $\xi_1\in T^{\sg}_m(M)$, $\xi_2\in T_{\sg(m)}(H_0)$, and
$\zeta\in T_n(N)$. Now, in order to verify (\ref{pluspre2}) given
(\ref{pluspre}), we examine the two possible cases allowed by
(\ref{decTMN}).
A dimension count shows that one can always choose $\al_h$ 
so as to satisfy (\ref{pluspre2}) on $T_{\sg(m)}(H_0)$. This is because
in a Lie groupoid one has \cite{Mac,NPL3}
$$
T_h(H)\simeq T^t_h(H)\oplus T_{t(h)}(H_0),
$$
and the condition (\ref{alhc}) constrains $\al_h$ only on
$T^t_h(H)$, leaving its value on $T_{t(h)}(H_0)$ free.
On the other hand, 
if (\ref{starstar}) holds, so that $(\dot{m}(0), (\dot{n}(0))$ lies in
$T_{(m,n)}^{\sg=\rh}(M\x N)$, and we assume (\ref{pluspre}), then
(\ref{pluspre2}) is satisfied for
any $\al_h$, as  one may choose
$\til{h}=\hat{h}=h$.  

Hence $\sim$ implies $\backsim$, and we have shown that these
equivalence relations coincide. Comparing (\ref{pluspre2})
with (\ref{defTGA}) and (\ref{defTGA2}), and using (\ref{sim1}),
it is  clear that (\ref{siso}) holds at the manifold
level. But it is almost
trivial that the identification we have made preserves the symplectic
structure, so that (\ref{siso}) is valid for symplectic manifolds as well.

Finally, we need to verify that the symplectomorphism (\ref{siso})
is compatible with the $A^*(G)$-$A^*(K)$ symplectic bimodule structure
that both sides have. This is, indeed, obvious from the explicit
structure of the pertinent Poisson maps. For example, 
denoting the appropriate Poisson map from
$T^*(M)^-\otc_{A^*(H)} T^*(N)^-$ to $A^*(G)$ by $\hat{J}_L^G$,
we have $\hat{J}_L^G([\theta_m,\et_n])=J^G_L(\theta_m)$, so that
\begeq
\left\langle \hat{J}_L^G([\theta_m,\et_n]),
\frac{d\gm(\lm)}{d\lm}_{|\lm=0}\right\rangle
= 
-\left\langle\theta_m,\frac{d\gm(\lm)\inv m}{d\lm}_{|\lm=0}\right\rangle.
\label{PI1}
\end{equation}
Here $[\theta_m,\et_n]$ is the equivalence class of $(\theta_m,\et_n)$
under either the $T^*(H)$ orbits or under the null foliation
with respect to the inclusion $T^*(M)^- *_{A^*(H)} T^*(N)^-\hookrightarrow
T^*(M)^-\x T^*(N)^-$; these coincide by (\ref{4lc}).

On the other hand, $\til{J}_L^G: T^*(M\otg_H N)^-\raw A^*(G)$ is given
by 
\begeq
 \left\langle\til{J}_L^G(\Theta_{[m,n]_H}),
\frac{d\gm(\lm)}{d\lm}_{|\lm=0}\right\rangle
= 
-\left\langle\Theta_{[m,n]_H},
\frac{d[\gm(\lm)\inv m,n]_H}{d\lm}_{|\lm=0}\right\rangle.\label{PI2}
\end{equation}

It is trivial from the explicit form of the isomorphism (\ref{siso}) 
described above that (\ref{PI1}) is duly transferred to (\ref{PI2}).

This completes the proof of the isomorphism between
(\ref{isolhs}) and (\ref{isorhs}), and therefore of
Theorem \ref{Thm2}.
\enp

Since Morita equivalence of  s-connected and s-simply connected 
Lie groupoids is isomorphism in
\LGb, and Morita equivalence of Poisson
manifolds is isomorphism in \Po,
we recover the known result \cite{NPLvE} that
the map $G\mapsto A^*(G)$ preserves Morita equivalence.


\begin{thebibliography}{99}
\itemsep=\smallskipamount
\bibitem{BDH}
Baillet, M., Denizeau, Y., Havet, J.-F. :  Indice d'une
esp\'{e}rance conditionnelle.   Compositio Math.\ \textbf{66},  
199--236 (1988)

\bibitem{Con0}  Connes, A.:
 Sur la th\'{e}orie non commutative de l'int\'{e}gration. 
 Lecture Notes in Math.\ \textbf{725},  19--143 (1979)

\bibitem{Con1} Connes, A.: 
A survey of foliations and operator algebras.
 Proc.\ Sympos.\ Pure Math.\ \textbf{38},   521--628  (1982)

\bibitem{Con}  Connes, A.: \textit{Noncommutative Geometry}.
San Diego: Academic Press,  1994

\bibitem{CT}  Connes, A., Takesaki, M.: The flow of weights on
factors of type III.  Tohoku Math.\ J.\  \textbf{29},  473--575 (1977).
Err.\ ibid.\  \textbf{30},  653--655 (1978)

\bibitem{CDW} 
Coste, A.,  Dazord, P.,  Weinstein, A.:
Groupoides symplectiques.  Publ.\ D\'{e}pt.\ Math.\ Univ. C.\
Bernard--Lyon I \textbf{2A}, 1--62 (1987)

\bibitem{Cou} 
Courant, T.J.: Dirac manifolds.  Trans.\ Amer.\ Math.\ Soc.\
\textbf{319},  631--661  (1990)

\bibitem{Daz}  Dazord,  P.: Groupo\"{\i}des symplectiques et troisi\`{e}me
 th\'{e}or\`{e}me de Lie ``non lin\'{e}aire''.  Lecture Notes in
Math.\ \textbf{1416},  39--74 (1990)

\bibitem{DS} Dazord, P., Sondaz, D.: Vari\'{e}t\'{e}s de Poisson --
alg\'{e}bro\"{\i}des de Lie.
S\'{e}minaire Sud-Rhodanien, Univ.\ Claude--Bernard, Lyon,  1--66 (1988)

\bibitem{FHM} Feldman,  J.,  Hahn,  P.,  Moore, C.C.: Orbit structure and
countable sections for actions of continuous groups.  Adv.\   Math.\  \textbf{28},
186--230 (1978)

\bibitem{Gre}  Green, P.:
 The local structure of twisted covariance algebras. Acta Math.\
 \textbf{140}, 191--250 (1978)

\bibitem{Hae} 
Haefliger, A.: Groupo\"{\i}des d'holonomie et classifiants. 
Ast\'{e}risque \textbf{116}, 70--97 (1984)

 \bibitem{Hah1} Hahn, P.: Haar measure for measure groupoids. 
Trans.\  Amer.\ Math.\ Soc.\ \textbf{242}, 1--33  (1978)

\bibitem{Hah2} Hahn, P.: 
The regular representations of measure
groupoids. Trans.\ Amer.\ Math.\ Soc.\ \textbf{242}, 35--72  (1978)

\bibitem{HS}
Hilsum, M.,   Skandalis, G.:
 Morphismes $K$-orient\'{e}s d'espaces de feuilles et fonctorialit\'{e}é
 en th\'{e}orie de
Kasparov (d'apr\'{e}s une conjecture d'A. Connes).  Ann.\ Sci.\ É\'{E}cole 
Norm.\ Sup.\ (4) \textbf{20},  325--390  (1987)

\bibitem{K1} 
Karasev, M.V.:
 The Maslov quantization conditions in higher cohomology and analogs
 of notions developed in Lie theory for canonical fibre bundles of
 symplectic manifolds.  I, II.   Selecta Math. Soviet.\ \textbf{8},
  213--234, 235--258 (1989)

 \bibitem{Kon} Kontsevich,  M.: Formal (non)commutative symplectic geometry.
In: \textit{The Gelfand Mathematical Seminars, 1990--1992}, pp.\ 173--187.
 Boston: Birkh\"{a}user,  1993

\bibitem{Krieger} Krieger,  W.: On ergodic flows and the isomorphism of factors.
Math. \ Ann.  \textbf{223}, 19--70 (1976)

\bibitem{KriMar} Krishnaprasad, P.S., Marsden, J.E.:
  Hamiltonian structure and stability for rigid bodies with
 flexible attachments. Arch.\ Rat.\ Mech.\ An.  {\bf 98},
 137--158 (1987)

\bibitem{Lance} 
Lance, E.C.: \textit{Hilbert
 $C^*$-Modules. A Toolkit for Operator Algebraists}.
 Cambridge:  Cambridge University Press, 1995

\bibitem{NPL3}  Landsman, N.P.: \textit{Mathematical Topics Between Classical
and Quantum Mechanics}.  New York: Springer, 1998

\bibitem{NPLCMP}  Landsman, N.P.:
  Lie groupoid $C^*$-algebras and Weyl quantization. 
 Commun.\  Math.\ Phys.\ \textbf{206},  367--381 (1999)

\bibitem{LR}  Landsman, N.P.,  Ramazan, B.:  
Quantization of Poisson algebras associated to Lie algebroids. In:
 Kaminker, J.,  Ramsay, A.,
 Renault, J.,  Weinstein, A. (eds.) 
\textit{Proc.\ Conf.\  on Groupoids in Physics, Analysis and
Geometry}.  Contemporary Mathematics,  to appear. E-print \texttt{math-ph/0001005}

\bibitem{NPLvE} Landsman, N.P.:  The Muhly--Renault--Williams theorem
for Lie groupoids and its classical counterpart. Lett.\ Math.\ Phys., to appear.
E-print  \texttt{math-ph/0008005}

\bibitem{NPLDR} Landsman,  N.P.: Bicategories of operator algebras
and Poisson manifolds. In: Longo, R. (ed.)
\textit{Mathematical Physics in Mathematics
and Physics.  Quantum and Operator Algebraic Aspects}. 
Fields Inst.\ Comm.,  to appear (2001).
E-print \texttt{math-ph/0008003}

\bibitem{NPLO} Landsman, N.P.:  Quantized reduction as a tensor product. In:
Landsman, N.P., Pflaum, M., Schlichenmaier, M. (eds.)
\textit{Quantization of Singular Symplectic Quotients}.  Basel:
 Birkh\"{a}user, 2001.
E-print \texttt{math-ph/0008004}

\bibitem{Mac}
Mackenzie,   K.C.H.: \textit{Lie Groupoids and Lie Algebroids in
Differential Geometry}.  Cambridge:  Cambridge University Press, 1987.

\bibitem{Mackey}   Mackey,  G.W.:
Ergodic theory and virtual groups.
Math.\ Ann.\ \textbf{166}, 187--207  (1966)

\bibitem{Moe} 
 Moerdijk,  I.:  Toposes and groupoids. 
 Lecture Notes in Math.\ \textbf{1348}, 280--298 (1988)

\bibitem{MM} 
 Moerdijk, I.,   Mr\v{c}un, J.: On integrability of infinitesimal
actions. Univ.\ of Utrecht preprint (April 2000)

\bibitem{Mol}  Molino, P.: \textit{Riemannian Foliations}.
  Basel: Birkh\"{a}user,  1988. 

\bibitem{MoS}  Moore, C.C.,  Schochet, C.:
\textit{Global Analysis on Foliated Spaces}.
New York: Springer,  1988

\bibitem{Mrc1}  Mr\v{c}un, J.: \textit{Stability and Invariants of
Hilsum--Skandalis Maps}. Ph.D.\ Thesis, Univ.\ of Utrecht (1996)

\bibitem{Mrc2} Mr\v{c}un, J.:
Functoriality of the bimodule associated to a Hilsum--Skandalis map.
 $K$-Theory \textbf{18},  235--253  (1999)

\bibitem{MRW} Muhly, P., Renault, J., 
  Williams, D.: Equivalence and 
isomorphism for groupoid \ca s. J.\ Operator Th.\ \textbf{17},
3--22  (1987)

\bibitem{Pra} Pradines, J.: Th\'{e}orie de Lie pour les
groupo\"{\i}des diff\'{e}rentiables. Relations entre
propri\'{e}t\'{e}s locales et globales.  C.\ R.\ Acad.\ Sc.\
Paris \textbf{A263}, 907--910  (1966)

\bibitem{Ram1}  Ramsay,  A.: Virtual groups and group actions.
Adv.\ Math.\ \textbf{6}, 253--322  (1971)

\bibitem{Ram2}  Ramsay,  A.: Topologies on measured groupoids.
J.\ Funct.\ Anal.\ \textbf{47}, 314--343 (1982)

\bibitem{Ren}
Renault, J.:  \textit{A Groupoid Approach to $C^*$-algebras}. 
Lecture Notes in Math.\ \textbf{793}. Berlin: Springer, 1980

\bibitem{Ren87} 
Renault, J.: Repr\'{e}sentation des produits crois\'{e}s
d'alg\`{e}bres de groupo\"{\i}des. J.\ Operator Theory \textbf{18},
 67--97  (1987)

\bibitem{Rie2} Rieffel, M.A.: 
Morita equivalence for $C^*$-algebras and
$W^*$-algebras.  J.\ Pure Appl.\ Alg.\ \textbf{5}, 51--96  (1974)

\bibitem{Sam} Samu\'{e}lid\`{e}s, M.: Mesures de Haar et $W\sp{*} $-couple d'un
groupo\"{\i}de mesur\'{e}. Bull.\ Soc.\ Math.\  France \textbf{106}, 261--278 (1978) 

\bibitem{Sau} 
Sauvageot, J.-L.:
Sur le produit tensoriel relatif d'espaces de Hilbert.
J.\ Operator Theory \textbf{9},  237--252  (1983)

\bibitem{Sau2} Sauvageot, J.-L.:
Image d'un homomorphisme et flot des poids
d'une relation d'\'{e}quivalence mesurée. Math.\  Scand. \textbf{42}, 71--100 
 (1978)

\bibitem{Sch}  Schweizer, J.: Crossed products by equivalence
bimodules.  Univ.\ T\"{u}bingen preprint (1999)

\bibitem{Sta} Stachura, P.:  Differential groupoids and
$C^*$-algebras.  E-print \texttt{math.QA/9905097}

\bibitem{SO} Stadler, M.O., O'Uchi, M.: 
Correspondence of groupoid \ca s.
J.\ Operator Th.  \textbf{42}, 103--119 (1999)


 
\bibitem{TS} Sutherland, C.E., Takesaki,  M.: Right inverse of the module
of approximately finite-dimensional factors of type III and approximately finite ergodic
principal measured groupoids.  Fields Inst.\  Commun. \textbf{20},
149--159  (1998)

\bibitem{Vai}  Vaisman, I.: \textit{Lectures on the Geometry of
Poisson Manifolds}.   Basel: Birkh\"{a}user,  1994

\bibitem{Was} 
Wassermann, A.: Operator algebras and conformal field
theory. III. Fusion of positive energy representations of ${\rm
LSU}(N)$ using bounded operators. Invent.\ Math.\ \textbf{133},
467--538 (1998)

\bibitem{W1} 
Weinstein, A.:  The
 local structure of Poisson manifolds.   J.\ Diff.\ Geom.  \textbf{
 18}, 523--557 (1983). Err.\ ibid.\ \textbf{22}, 255  (1985)

\bibitem{X1} Xu, P.:   Morita equivalent symplectic
 groupoids.   In: Dazord, P.,  Weinstein, A.  (eds.)
\textit{ Symplectic Geometry, Groupoids, and Integrable Systems}, pp.\ 291--311.
 New York:  Springer, 1991
  
 \bibitem{X2} Xu, P.:  Morita equivalence of Poisson
 manifolds.  Commun.\ Math.\ Phys.\ \textbf{142}, 493--509  (1991)

\end{thebibliography}
\end{document}